# A general classification of the replication dynamics with a unique fixed point in the interior of simplex $S_N$

Hongju (Daisy) Chen[1,2,3,4] Bin Yi[3] Zhanshan (Sam) Ma[3,4,5*]

[1]School of Mathematics and Statistics,
Guilin University of Technology,
Guilin, Guangxi, China

[2]Guangxi Colleges and Universities Key Laboratory of Applied Statistics,
Guilin, Guangxi, China

[3]Computational Biology and Medical Ecology Lab
Biostatistics and Image Genetics Lab
Kunming Institute of Zoology,
Chinese Academy of Sciences, Kunming, China

[4]Kunming College of Life Sciences,
University of Chinese Academy of Sciences, Kunming, China

[5]Faculty of Arts and Sciences
Harvard University
Cambridge, MA, 02138, USA
*For All Correspondence: ma@vandals.uidaho.edu

## Abstract

The replication dynamics (differential equation system) is the foundation of evolutionary game theory. When $n = 2$, there are four possible types of replication dynamics. When $n = 3$, there are 49 possible types of replication dynamics. However, when $n > 3$, the classification of replication dynamics has not been solved. In this article, the sufficient and necessary conditions of the replication dynamics equation with a unique fixed point in the interior of simplex $S_n$ ($IntS_n$) for $n \geq 2$ are presented. Furthermore, the different types of replication dynamics equations with a unique fixed point in $IntS_n$ is discussed.

# Introduction

Replicator dynamics have found wide application in computer science (Bloembergen et al., 2015; Tuyls et al., 2020; Ma & Krings 2011), economics (Sandholm, 2010), behavioural ecology and evolutionary biology (Maynard-Smith & Price 1972; Maynard-Smith 1982, McNamara & Leimar, 2020), and epidemiology (Nowak & May, 1994; Ma & Zhang 2025). A complete classification of the dynamics, especially for cases with a unique interior fixed point, is therefore of broad theoretical and practical interest.

Consider an evolutionary game of single population of many individuals with *N* strategy. Let A=$\left(a_{ij}\right)_{N\times N}$ be the payoff matrix of the game, where *N* is the total number of pure strategies of the game, $a_{ij} = \pi\left(R_i, R_j\right)$, is the payoff of a player (individual) with a pure strategy $R_i$ against another player (individual) with a pure strategy $R_j$.

The equation of replicator dynamics was first proposed by Taylor and Jonker (1978), and then named by Schuster and Sigmund (1983). It has become the most important equation in game dynamics (Cressman & Tao 2014). Suppose that $x_i$ is the frequency of choosing strategy $R_i$ in the population, then the replicator dynamics equations can be written as:

$$\begin{aligned}\dot{x}_1 &= x_1((Ax)_1 - xAx') \\ \dot{x}_2 &= x_2((Ax)_2 - xAx') \\ &\vdots \\ \dot{x}_N &= x_N((Ax)_N - xAx')\end{aligned} \quad (1)$$

where $x = (x_1, x_2, \dots, x_N) \in S_N$, $(Ax)_i = \sum_{j=1}^{N} x_j a_{ij}$, and $x'$ is the transpose of $x$ and $S_N$ represent a $N-1$ dimension simplex, that is

$$S_N = \{(x_1, x_2, \dots, x_N) | x_i \geq 0, \sum_{i=1}^{N} x_i = 1\}.$$

For $N \geq 2$, we divide the simplex $S_N$ into the following two parts:

(*i*) $\forall i \in \{1, 2, \dots, N\}, x_i > 0, \sum_{i=1}^{N} x_i = 1$, and $x = (x_1, x_2, \dots, x_N) \in S_N$ is the interior point of $S_N$. The set of all such points is called the interior of $S_N$ and denoted by $Int\mathrm{S_N}$, which means that each strategy does not go extinct.

(*ii*) $\exists i \in \{1, 2, \dots, N\}, x_i = 0$, and $x = (x_1, x_2, \dots, x_N) \in S_N$ is a boundary point of $S_N$. The set of all such points is called the boundary of $S_N$ and denoted by $Bd\mathrm{S_N}$, which means at least one of the strategies goes extinct.

The equations of replicated dynamics (1) can be divided into three cases (Hofbauer & Sigmund 1998; Nowak 2006):

(*i*) There is a unique fixed point in $Int\mathrm{S_N}$.

(*ii*) There are infinitely many fixed points in $Int\mathrm{S_N}$. There is now a manifold in $Int\mathrm{S_N}$, and these fixed points are stable, but not asymptotically stable.

(*iii*) There is no fixed point in $Int\mathrm{S_N}$, and all orbits will tend to $Bd\mathrm{S_N}$.

In evolutionary game theory, two equilibria are very important (Huttegger & Zollman 2013): One is (strictly) Nash equilibrium (NE) and the other is evolutionarily stable strategy (ESS) (Maynard & Price 1973). The fixed point of the replication dynamics equation is closely related to the NE. We summarize the conclusions as follows:

**Theorem A** (Hofbauer & Sigmund 2003; Cressman & Tao 2014), known as the Folk Theorem of Evolutionary Game Theory, states: (*i*) The point corresponding to the NE strategy must be the fixed point of Eqn. (1); (*ii*) The strategy corresponding to the fixed point of Lyapunov stable in Eqn. (1) must be NE; (*iii*) The strategy corresponding to the ω limit points of the inner orbitals of Eqn. (1) must be NE. (*iv*) The point corresponding to the strictly NE strategy must be the asymptotically stable fixed point of Eqn. (1).

Maynard & Price (1973) proposed the concept of ESS, and Taylor and Jonker (1978) proved that the fixed point corresponding to ESS is the asymptotically stable fixed point of the Eqn. (1). The study of the dynamic behavior (*such as* the fixed point and its stability) of the replication dynamics Eqn. (1) is essential for understanding the evolutionary game theory.

In fact, when $n = 2$, the payoff matrix is $A = \begin{pmatrix} a & b \\ c & d \end{pmatrix}$, then it has four possible types of replicator dynamics (Nowak 2006; Sigmund 2010), with the following figures:

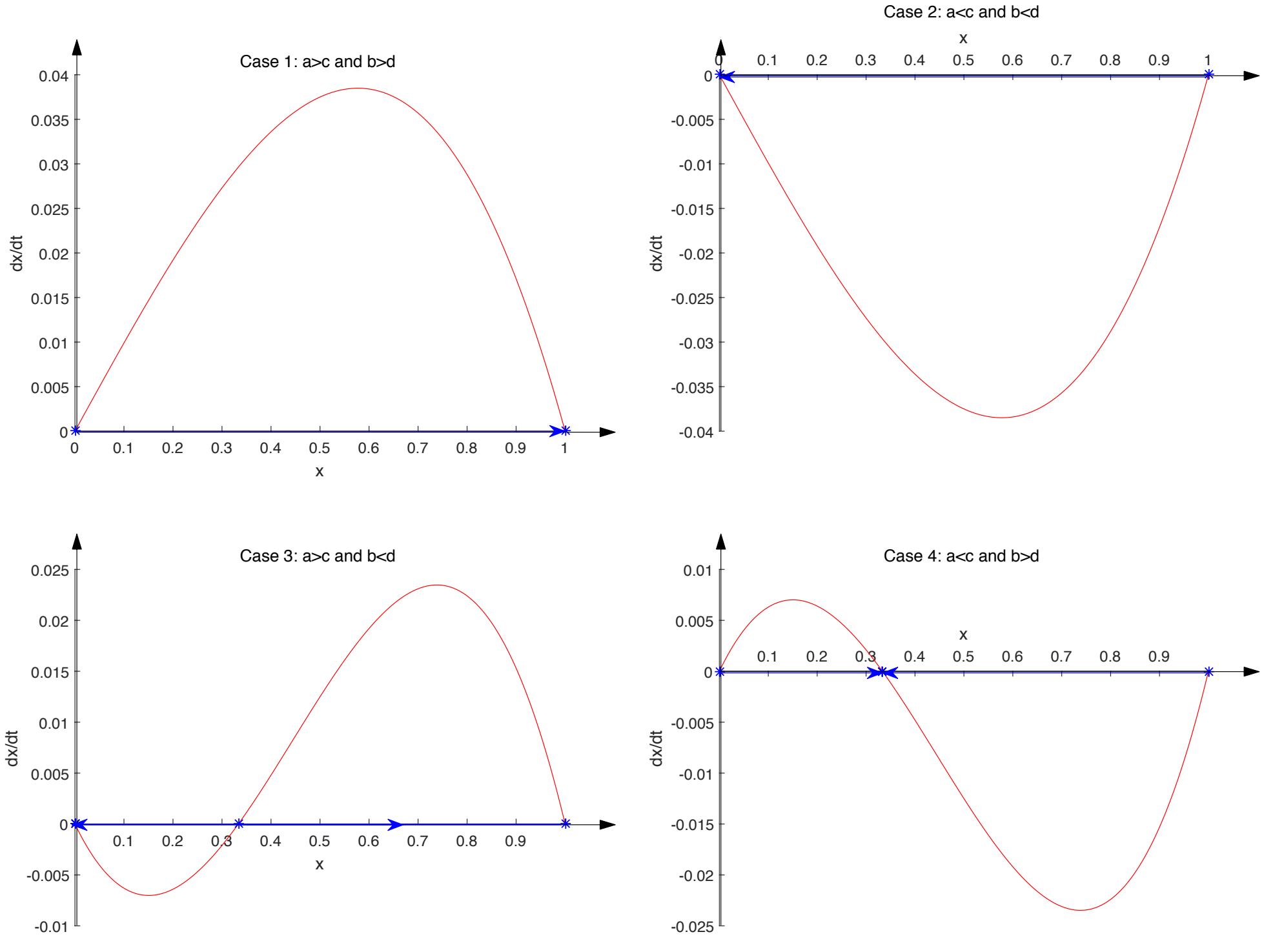

**Fig 1.** The figures of replicator dynamics when *n*=2.

Note:

(1) for case 1 and 2, there is no fixed point in $IntS_2$, and all the orbits tend to $BdS_2$.

(2) for case 3, there is only one fixed point for equation in $IntS_2$, and all the orbits (except for $x \equiv x^*$) tend to $BdS_2$.

(3) for case 4, there is only one fixed point for equation in $IntS_2$, and all the inner orbits tend to $x^*$.

When $n = 3$, the payoff matrix is $A = \begin{pmatrix} a_{11} & a_{12} & a_{13} \\ a_{21} & a_{22} & a_{23} \\ a_{31} & a_{32} & a_{33} \end{pmatrix}$, then it has 49 possible types of replicator dynamics (Bomze 1983, 1995). To the best of our knowledge, the classification of the replication dynamics equation for $n > 3$ has not been solved (Sigmund 2010). The objective of this study is to solve part of the problem.

# The main results

As introduced previously, the replication dynamics Eqn. (1) can be divided into three cases, *i.e.,* infinitely many fixed points in $IntS_N$, no fixed point in $IntS_N$, and a unique fixed point in $IntS_N$. This section includes the sufficient and necessary conditions of Eqn. (1), for the case that has a unique fixed point in $IntS_N$. The detailed proof for the sufficient and necessary condition is presented in this section.

**Theorem 1** Suppose that the payoff matrix of the single population with *N* strategies is $A = (a_{ij})_{N\times N}$, the sufficient and necessary conditions of the unique fixed point of replicator dynamics Eqn. (1) in $IntS_N$ are $A_{N\times i}A_{N\times(i+1)} > 0, (i = 1,2,\dots,N-1.)$, where $A_{N\times i}$ is the cofactor of the *N-th* row and the *i-th* column elements of the following determinant:

$$d = \begin{vmatrix} a_{11} - a_{21} & a_{12} - a_{22} & \cdots & a_{1N} - a_{2N} \\ a_{21} - a_{31} & a_{22} - a_{32} & \cdots & a_{2N} - a_{3N} \\ \vdots & \vdots & \ddots & \vdots \\ a_{(N-1)1} - a_{N1} & a_{(N-1)2} - a_{N2} & \cdots & a_{(N-1)N} - a_{NN} \\ 1 & 1 & \cdots & 1 \end{vmatrix}$$

and the fixed point of Eqn. (1) in $IntS_N$ is $x^* = (x_1{}^*, x_2{}^*, \dots, x_N{}^*)$, $x_i{}^* = \frac{A_{N\times i}}{d} = \frac{A_{N\times i}}{\sum_{i=1}^{N} A_{N\times i}}$.

Notice that: $x^* \in IntS_N$, thus $p^* = \sum_{i=1}^{N} x_i{}^* R_i$ is a Nash Equilibrium (NE), but it is not a strictly Nash Equilibrium (SNE).

***Proof*** Consider that these replicator dynamics equations are given by:

$$\begin{aligned} \dot{x}_1 &= x_1((Ax)_1 - xAx') \\ \dot{x}_2 &= x_2((Ax)_2 - xAx') \\ &\vdots \\ \dot{x}_N &= x_N((Ax)_N - xAx') \end{aligned} \tag{1}$$

where $x = (x_1, x_2, \dots, x_N) \in S_{N.}$ It is easy to see that the fixed points in Int$S_N$ if and only if

$$(Ax)_1 = (Ax)_2 = \cdots = (Ax)_N$$

$$x_1 + x_2 + \cdots + x_N = 1. \tag{2}$$

That is,

$$(Ax)_i - (Ax)_{i+1} = 0, \quad i = 1,2,\dots,\mathrm{N}-1$$
$$x_1 + x_2 + \cdots + x_N = 1$$

By expanding the above formula, we obtain:

$$\begin{cases} (a_{11} - a_{21})x_1 + (a_{12} - a_{22})x_2 + \cdots + (a_{1N} - a_{2N})x_N = 0 \\ (a_{21} - a_{31})x_1 + (a_{22} - a_{32})x_2 + \cdots + (a_{2N} - a_{3N})x_N = 0 \\ \vdots \\ (a_{(N-1)1} - a_{N1})x_1 + (a_{(N-1)2} - a_{N2})x_2 + \cdots + (a_{(N-1)N} - a_{NN})x_N = 0 \\ x_1 + x_2 + \cdots + x_N = 1 \end{cases} \tag{3}$$

This is an *N*-ary inhomogeneous linear equation for which the sufficient and necessary condition for a unique solution in $R^n$ is the determinant of coefficients

$$d = \begin{vmatrix} a_{11} - a_{21} & a_{12} - a_{22} & \cdots & a_{1N} - a_{2N} \\ a_{21} - a_{31} & a_{22} - a_{32} & \cdots & a_{2N} - a_{3N} \\ \vdots & \vdots & \ddots & \vdots \\ a_{(N-1)1} - a_{N1} & a_{(N-1)2} - a_{N2} & \cdots & a_{(N-1)N} - a_{NN} \\ 1 & 1 & \cdots & 1 \end{vmatrix} \neq 0$$

By expanding it on the N-*th* row, we discover that all the elements of the N-*th* row of the determinant $d$ are 1, and we get

$$d = \sum_{i=1}^{N} 1 \times A_{N\times i} = \sum_{i=1}^{N} A_{N\times i}$$

where $A_{N\times i}$ represent the cofactor of the *N-th* row and the i-*th* column element of the determinant *d*. That is,

$$A_{N\times i} = (-1)^{N+i} \times M_{N\times i},$$

where $M_{N\times i}$ represents the minor of the *N-th* row and the i-*th* column element of the determinant *d*.

By Cramer Rule, the unique solution of equation (2) can be written as:

$$x_i = \frac{d_i}{d}, i = 1,2, \dots, N.$$

Where $d_i$ $(i = 1, 2, \dots, N)$ is a new determinant, the *i-th* column element of the determinant $d$ is replaced by the vector $(0,0, \dots, 0,1)'$. If expand the determinant $d_i$ in terms of the *i-th* column, note that the first $N-1$ elements of the column are 0 and the last one is 1. Consequently, $d_i = A_{N\times i}, i = 1, 2, \dots, N$, then we get

$$x_i = \frac{A_{N\times i}}{d} = \frac{A_{N\times i}}{\sum_{i=1}^{N} A_{N\times i}}, i = 1,2, \dots, N.$$

However, for this solution to be in Int$S_N$, we must have

$$x_i = \frac{A_{N\times i}}{d} = \frac{A_{N\times i}}{\sum_{i=1}^{N} A_{N\times i}} > 0, i = 1,2, \dots, N.$$

This implies the numerator and the denominator have to be both positive or negative, *i.e.,* $A_{N\times i}A_{N\times(i+1)} > 0, i = 1, 2, \dots, N-1$. When the replicator dynamic equation (1) has a unique fixed point in Int$S_N$, and denote by

$$x^* = (x_1{}^*, x_2{}^*, \dots, x_N{}^*), x_i{}^* = \frac{A_{N\times i}}{d} = \frac{A_{N\times i}}{\sum_{i=1}^{N} A_{N\times i}} > 0, i = 1,2, \dots, N.$$

Denote $\boldsymbol{p}^* = \sum_{i=1}^{\mathrm{N}} x_i{}^* \mathrm{R}_i$, then by the above proof, we have

$$(Ax^*)_i = x^* A x^{*\prime}, \quad i = 1,2, \dots, \mathrm{N}$$

that is

$$\forall i \in \{1,2, \dots, \mathrm{N}\}, \pi(\mathrm{R}_i, \boldsymbol{p}^*) = \pi(\boldsymbol{p}^*, \boldsymbol{p}^*).$$

So, for any strategy

$$\boldsymbol{p} = \sum_{i=1}^{\mathrm{N}} x_i \mathrm{R}_i\ , \quad x = (x_1, x_2, \dots, x_N) \in S_N$$

we have $\pi(\boldsymbol{p}, \boldsymbol{p}^*) = \pi(\boldsymbol{p}^*, \boldsymbol{p}^*)$. Therefore, $\boldsymbol{p}^* = \sum_{i=1}^{\mathrm{N}} x_i{}^* \mathrm{R}_i$ is a Nash Equilibrium.

Theorem 1 has been proved.

The following corollary can be drawn from the above Theorem 1:

**Corollary 1** If the determinant $d \neq 0$ in Theorem 1 and $\exists i \in \{1,2, \dots, N-1\}$, such that $A_{N\times i}A_{N\times(i+1)} \leq 0$, there is no solution for Eqn. (1) in $Int\mathrm{S_N}$, then every orbit tends to $Bd\mathrm{S_N}$.

***Proof*** Since the determinate $d \neq 0$ of theorem 1 and $\exists i \in \{1,2,\dots,N-1\}$, such that $A_{N\times i}A_{N\times(i+1)} \leq 0$. By the proof of theorem 1, there is only one solution in $R^n$ , but not in $\mathrm{Int}S_N$. Therefore, there is no solution of equation (1) in $\mathrm{Int}S_N$, and every orbit tends to $BdS_N$.

As shown in figure 1, for case 3 and case 4, there is only one solution $x^*$ in $\mathrm{Int}S_2$. For case 3, all the orbits (except for $x \equiv x^*$) tend to $BdS_2$, and for case 4, all inner orbits tend to $x^*$. The general case for $N \geq 2$, we obtain the following results:

**Theorem 2** Suppose that $A = (a_{\mathrm{ij}})_{N\times N}$ is the payoff matrix of a single population with *N* strategies that satisfied the condition of Theorem 1, *B* is a *N-1* symmetric square matrix with $b_{ij} = -(a_{ij} + a_{ji} - a_{Ni} - a_{iN} - a_{Nj} - a_{jN} + 2a_{NN}), 1 \leq i,j \leq N-1$, there are: (*i*) $\boldsymbol{p}^*$is an ESS if and only if $B$ is a positive definite matrix, particularly, $x = x^*$ is globally asymptotically stable and all inner orbits converge to $x^*$. $\boldsymbol{p}^*$ is a NE, but it is not an ESS if and only if $B$ is not a positive definite matrix. (*ii*) If *B* is a positive semi-definite matrix, then $x^*$ is Lyapunov stable, particularly, if $B = 0$, then $\prod_{i=1}^{N} {x_i}^{{x_i}^*} \equiv c$. (*iii*) if *B* is a negative definite matrix, then $x^*$ is unstable and every orbit tends to $Bd\mathrm{S_N}$.

To prove the Theorem 2, we first give the following two lemmas:

**Lemma 1** (Hofbauer *et al.* 1998, P64)**:** The strategy $\boldsymbol{p}$ is an ESS in $\mathrm{Int}S_N$ if and only if

$$pAq' = \pi(\boldsymbol{p},\boldsymbol{q}) > \pi(\boldsymbol{q},\boldsymbol{q}) = qAq'$$

For all strategy $\boldsymbol{q} \neq \boldsymbol{p}$ and $q \in S_N$.

**Lemma 2** Suppose that

$$f(x) = f(x_1,\dots,x_n) = (x_1,\dots,x_n)\begin{pmatrix} a_{11} & \cdots & a_{1n} \\ \vdots & \ddots & \vdots \\ a_{n1} & \cdots & a_{nn} \end{pmatrix}\begin{pmatrix} x_1 \\ \vdots \\ x_n \end{pmatrix},$$

where $x = (x_1, \dots, x_n) \in R^n$, $A = \begin{pmatrix} a_{11} & \cdots & a_{1n} \\ \vdots & \ddots & \vdots \\ a_{n1} & \cdots & a_{nn} \end{pmatrix}$ is a symmetric matrix and $\delta$ is a positive number, then $f(x) > 0$ in $U^o(0,\delta) = \left\{x \middle| 0 < \|x\| = \sqrt{{x_1}^2 + \cdots + {x_n}^2} < \delta\right\}$ if and only if the matrix $A$ is a positive definite matrix.

***Proof*** If $A = \begin{pmatrix} a_{11} & \cdots & a_{1n} \\ \vdots & \ddots & \vdots \\ a_{n1} & \cdots & a_{nn} \end{pmatrix}$ is a positive definite matrix, then $f(x) > 0$ holds for all $x \neq 0$. Naturally, $f(x) > 0\ in\ U^o(0,\delta) = \left\{x \middle| 0 < \|x\| = \sqrt{{x_1}^2 + \cdots + {x_n}^2} < \delta\right\}$.

Assume $f(x) > 0\ in\ U^o(0,\delta) = \left\{x \middle| 0 < \|x\| = \sqrt{{x_1}^2 + \cdots + {x_n}^2} < \delta\right\}$. By proof by contradiction, we suppose $A = \begin{pmatrix} a_{11} & \cdots & a_{1n} \\ \vdots & \ddots & \vdots \\ a_{n1} & \cdots & a_{nn} \end{pmatrix}$ is not a positive definite matrix, then there must exist $x^* = ({x_1}^*, \dots, {x_n}^*) \in R^n, x^* \neq 0$ such that $f(x^*) \leq 0$. Thus, there exist a large number $M > 0$ such that $\left\|\frac{x^*}{M}\right\| < \delta$, *i.e.*, $\frac{x^*}{M} \in U^o(0,\delta)$. However, $f\left(\frac{x^*}{M}\right) = \frac{1}{M^2} f(x^*) \leq 0$. This is a contradiction. Thus, $A = \begin{pmatrix} a_{11} & \cdots & a_{1n} \\ \vdots & \ddots & \vdots \\ a_{n1} & \cdots & a_{nn} \end{pmatrix}$ is a positive definite matrix. This completes the proof.

The proof of theorem 2 is presented based on the above two Lemmas.

***Proof:***

1) Let $x = (x_1, x_2, \dots, x_N) \in S_N$, $\boldsymbol{q} = \sum_{i=1}^{\mathrm{N}} x_i \mathrm{R}_i$, by lemma 1, the strategy $\boldsymbol{p}^*$ is an ESS if and only if $x^* \mathrm{A} x' = \pi(\boldsymbol{p}^*, \boldsymbol{q}) > \pi(\boldsymbol{q}, \boldsymbol{q}) = x \mathrm{A} x'$ holds for all $x \in S_N, x \neq x^*$.

Let $f(x) = f(x_1, x_2, \dots, x_N) = x^* \mathrm{A} x' - x \mathrm{A} x'$, it is easy to see that $f(x^*) = 0$, where $f(x)$ is an *N*-ary quadratic function.

So, the strategy $\boldsymbol{p}^*$ is an ESS if and only if

$$f(x) = f(x_1, x_2, \dots, x_N) = x^*\mathrm{A}x' - x\mathrm{A}x' > 0 = f(x^*)$$

holds for all $x \in S_N, x \neq x^*$.

That is, strategy $\boldsymbol{p}^*$is an ESS if and only if $x^*$ is the strict minimum point of $f(x)$ on the $x = \{(x_1, x_2, \dots, x_N) \big| x_i \geq 0, \sum_{i=1}^{N} x_i = 1\}$, where $f(x) = f(x_1, x_2, \dots, x_N) = x^*\mathrm{A}x' - x\mathrm{A}x'$.

Note that $x_N = 1 - x_1 - \cdots - x_{N-1}$, $x_N{}^* = 1 - x_1{}^* - \cdots - x_{N-1}{}^*$, thereby,

$$f(x) = f(x_1, x_2, \dots, x_N) = x^*\mathrm{A}x' - x\mathrm{A}x'$$

$$= (x_1{}^*, \dots, x_{N-1}{}^*, 1 - x_1{}^* - \cdots - x_{N-1}{}^*)A\begin{pmatrix} x_1 \\ \vdots \\ x_{N-1} \\ 1 - x_1 - \cdots - x_{N-1} \end{pmatrix}$$

$$-(x_1, \dots, x_{N-1}, 1 - x_1 - \cdots - x_{N-1})A\begin{pmatrix} x_1 \\ \vdots \\ x_{N-1} \\ 1 - x_1 - \cdots - x_{N-1} \end{pmatrix}.$$

It is a quadratic function of the $N$-1 element with respect to $x_1, \dots, x_{\mathrm{N}-1}$, which is denoted by $g(x_1, x_2, \dots, x_{N-1}) = f(x_1, x_2, \dots, x_N)$.

Note that, $x_i{}^* = \frac{A_{N\times i}}{d} = \frac{A_{N\times i}}{\sum_{i=1}^{N} A_{N\times i}} > 0$ holds for any $i \in \{1, 2, \dots, N\}$.

Set $\delta = \min\left\{\frac{x_i{}^*}{N}, 1 \leq i \leq N\right\}$, we have $\delta > 0$.

Denote $y^* = (x_1{}^*, \dots, x_{N-1}{}^*) \in \mathrm{R}^{N-1}, y = (x_1, x_2, \dots, x_{N-1}) \in \mathrm{R}^{N-1}$ ; then we have $g(y) = g(x_1, x_2, \dots, x_{N-1})$ is defined on the set $U(y^*, \delta) = \left\{y \middle| \|y - y^*\| = \sqrt{\sum_{i=1}^{N-1} |x_i - x_i{}^*|^2} < \delta\right\} \subset R^{N-1}$.In other words, when $y = (x_1, \dots, x_{N-1}) \in U(y^*, \delta)$, and let $x_N = 1 - x_1 - \cdots - x_{N-1}$, then we have $(x_1, \dots, x_{N-1}, x_N) \in S_N$ hold.

However, $g(x_1, x_2, \dots, x_{N-1}) = f(x_1, x_2, \dots, x_N)$

$$= (x_1{}^*, \dots, x_{N-1}{}^*, x_N{}^*) A \begin{pmatrix} x_1 \\ \vdots \\ x_{N-1} \\ x_N \end{pmatrix} - (x_1, \dots, x_{N-1}, x_N) A \begin{pmatrix} x_1 \\ \vdots \\ x_{N-1} \\ x_N \end{pmatrix}$$

$$= \sum_{i=1}^{N} x_i{}^* \, (Ax)_i - \sum_{i=1}^{N} x_i \, (Ax)_i$$

where $x_N = 1 - x_1 - \cdots - x_{N-1}$, $x_N{}^* = 1 - x_1{}^* - \cdots - x_{N-1}{}^*$, $(Ax)_i = \sum_{j=1}^{N} x_j a_{ij}$. Note that it is an $N-1$-ary quadratic function in terms of $x_1, \dots, x_{N-1}$; therefore, there are continuous second-order partial derivatives, and all the second-order partial derivatives are constant. We next prove that $y^* = (x_1{}^*, \dots, x_{N-1}{}^*)$is a stationary point of $g(x_1, x_2, \dots, x_{N-1})$. By calculation, we have:

$$\frac{\partial g}{\partial x_1} = \sum_{i=1}^{N} x_i{}^* \frac{\partial (Ax)_i}{\partial x_1} - \sum_{i=1}^{N} \left(\frac{\partial x_i}{\partial x_1} (Ax)_i + x_i \frac{\partial (Ax)_i}{\partial x_1}\right)$$

$$= \sum_{i=1}^{N} x_i{}^* \, (a_{i1} - a_{iN}) - ((Ax)_1 - (Ax)_N) - \sum_{i=1}^{N} x_i (a_{i1} - a_{iN})$$

$$= \sum_{i=1}^{N} (x_i{}^* - x_i) \, (a_{i1} - a_{iN}) - ((Ax)_1 - (Ax)_N)$$

It is worth noting that when $x_i = x_i{}^* (i = 1,2, \dots, N-1)$, there must be

$$x_N = 1 - x_1 - \cdots - x_{N-1} = 1 - x_1{}^* - \cdots - x_{N-1}{}^* = x_N{}^*.$$

Note that $x^* = (x_1{}^*, x_2{}^*, \dots, x_N{}^*)$ is a solution of equation (2), thus we have $(\mathrm{A}x^*)_1 = (\mathrm{A}x^*)_\mathrm{N}$.

Hence, $\frac{\partial g}{\partial x_1}|_{(x_1{}^*, \dots, x_{N-1}{}^*)} = 0$ holds. Similarly,

$$\frac{\partial g}{\partial x_i}|_{(x_1{}^*, \dots, x_{N-1}{}^*)} = 0 (i = 2,3, \dots, N-1).$$

Thus, $y^* = (x_1{}^*, \dots, x_{N-1}{}^*)$ is a stationary point of $g(x_1, x_2, \dots, x_{N-1})$.

Because $g(x_1, x_2, \dots, x_{N-1})$is an $N$-1-ary quadratic function with respect to $x_1, \dots, x_{N-1}$, which has continuous second-order partial derivatives, third-order partial derivatives and above on $U(\mathrm{y}^*, \delta) \subset \mathrm{R}^{\mathrm{N}-1}$ are equal to 0, and $y^* = (x_1{}^*, \dots, x_{N-1}{}^*)$ is a stationary point of $g(x_1, x_2, \dots, x_{N-1})$. Therefore, by Taylor Formula expansion of the multivariate function, we have:

$$g(x_1,\dots,x_{N-1}) = g(x_1{}^*,\dots,x_{N-1}{}^*) + \sum_{i=1}^{N-1}\frac{\partial g}{\partial x_i}|_{(x_1{}^*,\dots,x_{N-1}{}^*)} \times (x_i - x_i{}^*)$$

$$+\frac{1}{2}(x_1 - x_1{}^*,\dots,x_{N-1} - x_{N-1}{}^*)Hg(y^*)\begin{pmatrix} x_1 - x_1{}^* \\ \vdots \\ x_{N-1} - x_{N-1}{}^* \end{pmatrix}$$

$$= \tfrac{1}{2}(x_1 - x_1{}^*,\dots,x_{N-1} - x_{N-1}{}^*)Hg(y^*)\begin{pmatrix} x_1 - x_1{}^* \\ \vdots \\ x_{N-1} - x_{N-1}{}^* \end{pmatrix} \qquad (4)$$

where

$$\text{Hg}(y^*) = (b_{ij})_{(N-1)(N-1)}, b_{ij} = \frac{\partial^2 g}{\partial x_i \partial x_j}|_{(x_1{}^*,\dots,x_{N-1}{}^*)} = \frac{\partial^2 g}{\partial x_j \partial x_i}|_{(x_1{}^*,\dots,x_{N-1}{}^*)} = b_{ji},$$

$$1 \le i,j \le N-1.$$

It is a Hessian matrix of $g(x_1, x_2, \dots, x_{N-1})$ at the point $(x_1{}^*,\dots,x_{N-1}{}^*)$ and a symmetric square matrix. By calculation, we get

$$b_{ij} = -(a_{ij} + a_{ji} - a_{Ni} - a_{iN} - a_{Nj} - a_{jN} + 2a_{NN}), 1 \le i,j \le N-1.$$

If strategy $\boldsymbol{p}^*$ is an ESS, then $y^*$is a strict minimum point of

$$g(y) = g(x_1,\dots,x_{N-1}) = x^*\text{A}x' - x\text{A}x'$$

on the set $x = \{(x_1, x_2, \dots, x_N)|x_i \ge 0, \sum_{i=1}^{N} x_i = 1\}$. Thus, there must be a strict minimum point of $g(y)$ on $U(y^*,\delta) \subset R^{N-1}$. By lemma 2, $\text{Hg}(y^*)$ is a positive definite matrix.

If $\text{Hg}(y^*)$ is a positive definite matrix, by Equation (4), $g(x_1,\dots,x_{N-1}) > 0$ holds for all $(x_1, x_2, \dots, x_{N-1}) \ne (x_1{}^*,\dots,x_{N-1}{}^*)$. Consequently, $y^*$ is the strict minimum point of $g(y) = g(x_1,\dots,x_{N-1}) = x^*\text{A}x' - x\text{A}x'$ for any $x = (x_1, x_2, \dots, x_N) \in S_N$. Hence, the strategy $\boldsymbol{p}^*$ is an ESS.

2) Set

$$V(x) = \prod_{i=1}^{N} x_i{}^{*x_i{}^*} - \prod_{i=1}^{N} x_i{}^{x_i{}^*} \quad x = (x_1, x_2, \dots, x_N) \in S_N$$

then, $V(x) \ge 0$, and $V(x) = 0 \Leftrightarrow x = x^* = (x_1{}^*, x_2{}^*, \dots, x_N{}^*)$.

Thus, by the derivation process of 1) and calculation, we have:

$$\dot{V(x)} = \frac{dV(x)}{dt} = -(x^*\mathrm{A}x' \ - x\mathrm{A}x' \ )\times \prod_{i=1}^{N} x_i^{x_i^*}$$

$$= -\frac{1}{2}(x_1 - x_1^*, \dots, x_{N-1} - x_{N-1}^*)B\begin{pmatrix} x_1 - x_1^* \\ \vdots \\ x_{N-1} - x_{N-1}^* \end{pmatrix}\times \prod_{i=1}^{N} x_i^{x_i^*}$$

If $B$ is a positive semi-definite matrix, then $\forall x = (x_1, x_2, \dots, x_N) \in S_N, x \neq x^*$, we have

$$\dot{V(x)} \leq 0$$

thus, $x^*$ is a fixed point of Lyapunov stability of the equation (1). Particularly, if $B = 0$, then $\dot{V(x)} \equiv 0$. Thus, $\prod_{i=1}^{N} x_i^{x_i^*} \equiv c$, where $c$ is determined by its initial value.

3) if B is a negative definite matrix, then $\forall x = (x_1, x_2, \dots, x_N) \in S_N, x \neq x^*$, we have

$$\dot{V(x)} = -\frac{1}{2}(x_1 - x_1^*, \dots, x_{N-1} - x_{N-1}^*)B\begin{pmatrix} x_1 - x_1^* \\ \vdots \\ x_{N-1} - x_{N-1}^* \end{pmatrix}\times \prod_{i=1}^{N} x_i^{x_i^*} > 0.$$

That is, any orbits except $x \equiv x^*$, we obtain

$$V(x) = \prod_{i=1}^{N} {x_i^*}^{x_i^*} - \prod_{i=1}^{N} x_i^{x_i^*}$$

is strictly monotonically increasing. It is note that

$$\max_{x \in S_N} V(x) = \sup_{x \in Int S_N} V(x) = \max_{x \in BdS_N} V(x) = \prod_{i=1}^{N} {x_i^*}^{x_i^*},$$

thus, every orbital except $x \equiv x^*$ tends to $BdS_N$. Theorem 2 has been proved.

Note that there are only five possibilities for matrix B, including a positive definite matrix, a positive semi-definite matrix, a negative definite matrix, a negative semi-definite matrix and an indefinite matrix. Based on the above theorem, $\boldsymbol{p}^*$ is a NE but not an ESS if and only if $B$ is not a positive definite matrix; if $B$ is a positive semi-definite matrix, then $x \equiv x^*$ is Lyapunov stable; if $B$ is a negative definite matrix, then except for $x \equiv x^*$, every orbit tends to $BdS_N$. In the next section, some examples are presented to illustrate that: if $B$ is an indefinite matrix or a negative semi-definite matrix except for $B = 0$, then some orbitals may tend to $Bd\mathrm{S_N}$ and some orbitals may tend to $x^*$.

# Discussion

In this section, we demonstrate that some existing results in the literature of the replication dynamics are the special cases of the results (theorems) presented in this article.

## Hawk-dove game (HD) with only one ESS in $IntS_N$ ($N \geq 2$)

We first discuss the well familiar hawk-dove game (Maynard-Smith & Price 1973; Maynard-Smith 1983). Conflicts between animals of same species over food, territory, mates, *etc.,* in the animal kingdom, can be brutal and even deadly. However, under normal circumstances, conflicts between animals do not escalate, which is called conventional struggle by animal behaviorists. For example, in many species of snakes, males fight each other by wrestling rather than using their fangs. In growling contests, the males move parallel to each other and lock their antlers head forward, although the damage from antler attacks can be fatal, but this is rare. To date, most biologists have accepted the following view of the situation: conventional struggles are good for species, and fights that result in serious damages are bad for species. It turns out that in nature such conventional struggles are in great majority. However, this does not explain why sometimes it escalates the fight by breaking the rules (that is, inflicting serious damage on its opponent). In such exceptions, it can beat the competitors and achieves the final victory. When the mutation for breaking the rule occurs, it may have a higher gene replication rate than any other members! (Hofbauer *et al.* 1998; Nowak 2006).

Maynard-Smith and Price (1973) first explained the conventional struggle from the perspective of individual choice. Now suppose that there are two strategies, the hawk strategy (H) and the dove strategy (D). The hawk strategy is to escalate the fights, while the dove strategy is the opposite (*such as* retreating when the other side tries to escalate the fight). The payoff of the winner is *b,* and the payoff of the loser is *c* ($b<c$). If both players choose the hawk strategy, then the fight will escalate, and their final expected payoff is $\frac{b-c}{2}$. If one takes the hawk strategy and the other takes the dove strategy, then the former wins and gets the payoff is *b*, and the latter retreats and gets the payoff is 0. If both players take the dove strategy, neither player loses anything, and each player has a fifty-fifty chance of winning, then their expected payoff is $\frac{b}{2}$. Therefore, the payoff matrix of the Hawk-dove game can be derived as follows:

$$A = \begin{pmatrix} \frac{b-c}{2} & b \\ 0 & \frac{b}{2} \end{pmatrix}$$

where $R_1 = H,\ R_2 = D,$ are the hawk strategy and dove strategy respectively. Assume that the frequency of hawk strategy selection in this population be $x$; then the frequency of dove strategy selection is $1 - x$. According to replicator dynamics, there is:

$$\dot{x} = x(1-x)(\frac{b}{2} - \frac{c}{2}x).$$

Let $\dot{x} = 0$, and three fixed points can be derived as: $x = 0, 1, \frac{b}{c},$ where $x = 0, 1$ is unstable, and $x = \frac{b}{c}$ is globally stable.

In fact, it can be further shown that $\boldsymbol{p} = \frac{b}{c}H + \left(1 - \frac{b}{c}\right)D$ is an evolutionarily stable strategy (ESS) (Maynard-Smith 1982; Hofbauer *et al.* 1998; Nowak 2006). Particularly, when $c \gg b$, it indicates that in the evolutionarily stable strategy, most individuals in the population will choose the dove strategy, which adequately explains the phenomenon of conventional struggles (*i.e.,* no escalation). Thus, the ESS has two characteristics:

**Characteristics 1** The strategy $\boldsymbol{p} = \frac{b}{c}H + \left(1 - \frac{b}{c}\right)D$ is a mixed strategy, where $p = (\frac{b}{c}, 1 - \frac{b}{c}) \in IntS_2$. It is an evolutionarily stable strategy (ESS) but is not a strictly Nash Equilibrium (SNE).

**Characteristics 2** From the point of view of evolutionary dynamics, this ESS is the only evolutionary endpoint.

This naturally raises the following questions:

**Question 1** What are the sufficient and necessary conditions for any single population with *N* strategy game to have ESS in $IntS_N$ $(N \geq 2)$?

In fact, when *N*=2, the answer is as follows (Maynard-Smith 1982; Hofbauer *et al,* 1998): Suppose the payoff matrix $A = \begin{pmatrix} a_{11} & a_{12} \\ a_{21} & a_{22} \end{pmatrix}$, then the sufficient and necessary conditions for ESS in $Int\mathrm{S}_2$ are that $a_{11} < a_{21}$ and $a_{22} < a_{12}$.

By Theorems 1 and 2, the sufficient and necessary conditions for the equations (1) in $Int\mathrm{S}_\mathrm{N}$ $(N \geq 2)$ to have ESS are:

$(i)$ $A_{N\times i}A_{N\times(i+1)} > 0, (i = 1,2,\dots,N-1.)$, where $A_{N\times i}$ is the cofactor of the *N-th* row and the *i-th* column elements of the following determinant:

$$d = \begin{vmatrix} a_{11} - a_{21} & a_{12} - a_{22} & \cdots & a_{1N} - a_{2N} \\ a_{21} - a_{31} & a_{22} - a_{32} & \cdots & a_{2N} - a_{3N} \\ \vdots & \vdots & \ddots & \vdots \\ a_{(N-1)1} - a_{N1} & a_{(N-1)2} - a_{N2} & \cdots & a_{(N-1)N} - a_{NN} \\ 1 & 1 & \cdots & 1 \end{vmatrix}.$$

$(ii)$ $B = (b_{ij})$ is a symmetric square matrix of order *N*-1 and a positive definite matrix, where $b_{ij} = -(a_{ij} + a_{ji} - a_{Ni} - a_{iN} - a_{Nj} - a_{jN} + 2a_{NN}), 1 \leq i,j \leq N-1$.

In summary, this completely solves Question 1. Particularly, if these two conditions are true for matrix A, then $p^* = \sum_{i=1}^{N} x_i^* R_i$ is the only endpoint of evolution, where $x_i^* = \frac{A_{N\times i}}{d} = \frac{A_{N\times i}}{\sum_{i=1}^{N} A_{N\times i}}$.

## Rock- Scissors-Paper (RSP) game with *N*=3 strategies

Now let us discuss the *N*=3 case. Assume that $A = \begin{pmatrix} a_{11} & a_{12} & a_{13} \\ a_{21} & a_{22} & a_{23} \\ a_{31} & a_{32} & a_{33} \end{pmatrix}$, without loss of generality, the elements in the diagonal are 0 (Nowak 2006, P58), *i.e.,*

$$A = \begin{pmatrix} 0 & a_{12} & a_{13} \\ a_{21} & 0 & a_{23} \\ a_{31} & a_{32} & 0 \end{pmatrix}.$$

By Theorem 1, it can be derived that the sufficient and necessary conditions for the Eqn. (1) to have a unique fixed point in $Int\mathrm{S}_3$ are that the following conditions be satisfied:

$$A_{31} > 0,\ A_{32} > 0,\ A_{32} > 0, \text{ or } A_{31} < 0, A_{32} < 0, A_{32} < 0$$

where

$$A_{31} = \begin{vmatrix} a_{12} & a_{13} - a_{23} \\ -a_{32} & a_{23} \end{vmatrix},\ A_{32} = -\begin{vmatrix} -a_{21} & a_{13} - a_{23} \\ a_{21} - a_{31} & a_{23} \end{vmatrix},$$

$$A_{33} = \begin{vmatrix} -a_{21} & a_{12} \\ a_{21} - a_{31} & -a_{32} \end{vmatrix}$$

In this case, the Eqn. (1) has a unique fixed point in $IntS_3$ and denoted by $x^* = (x_1{}^*, x_2{}^*, x_3{}^*)$, where $x_i{}^* = \frac{A_{3i}}{A_{31}+A_{32}+A_{33}}$.

Thus,

$$p^* = \sum_{i=1}^{3} \frac{A_{3i}}{A_{31} + A_{32} + A_{33}} R_i$$

is a Nash Equilibrium (NE)., but it's not a strictly Nash Equilibrium (SNE).

By Theorem 2, set

$$B = \begin{pmatrix} 2(a_{13} + a_{31}) & a_{13} + a_{31} + a_{23} + a_{32} - a_{12} - a_{21} \\ a_{13} + a_{31} + a_{23} + a_{32} - a_{12} - a_{21} & 2(a_{32} + a_{23}) \end{pmatrix},$$

thus $p^*$ is an ESS if and only if $B$ is positive definite matrix. $B$ is a positive definite matrix if and only if all leading principal minors are greater than 0. That is, when

$$2(a_{13} + a_{31}) > 0,$$

and

$$\begin{vmatrix} 2(a_{13} + a_{31}) & a_{13} + a_{31} + a_{23} + a_{32} - a_{12} - a_{21} \\ a_{13} + a_{31} + a_{23} + a_{32} - a_{12} - a_{21} & 2(a_{32} + a_{23}) \end{vmatrix} > 0,$$

then $p^*$ is an ESS. In other words, if $B$ does not satisfy these conditions, then $\boldsymbol{p}^*$ is a Nash equilibrium, but not an ESS.

When $B$ is a positive semi-definite matrix ($B$ is a positive semi-definite matrix if and only if all principal minors of $B$ are non-negative), that is, when $2(a_{13} + a_{31}) \geq 0, 2(a_{32} + a_{23}) \geq 0$, and

$$\begin{vmatrix} 2(a_{13} + a_{31}) & a_{13} + a_{31} + a_{23} + a_{32} - a_{12} - a_{21} \\ a_{13} + a_{31} + a_{23} + a_{32} - a_{12} - a_{21} & 2(a_{32} + a_{23}) \end{vmatrix} \geq 0,$$

the fixed point $x^*$ is Lyapunov stable. Particularly, if $B = 0$, then $\prod_{i=1}^{N} x_i{}^{x_i{}^*} \equiv c$.

When $B$ is a negative definite matrix ($B$ is negative definite matrix if and only if all the odd-order leading principal minors of $B$ are less than 0 and all the even-order leading principal minors of $B$ are greater than 0), that is,

when $2(a_{13} + a_{31}) < 0$ and

$$\begin{vmatrix} 2(a_{13}+a_{31}) & a_{13}+a_{31}+a_{23}+a_{32}-a_{12}-a_{21} \\ a_{13}+a_{31}+a_{23}+a_{32}-a_{12}-a_{21} & 2(a_{32}+a_{23}) \end{vmatrix} > 0,$$

all orbitals except $x \equiv x^*$ tends to $BdS_N$.

Now let us consider the Rock-Scissors-Paper game with matrix:

$$A = \begin{pmatrix} 0 & a & -1 \\ -1 & 0 & a \\ a & -1 & 0 \end{pmatrix}$$

where $a > 0,$ Eqn. (1) has a unique fixed point in $Int\mathrm{S}_3,$ that is,

$$x^* = \left(\frac{1}{3}, \frac{1}{3}, \frac{1}{3}\right) \text{ and } B = \begin{pmatrix} 2(a-1) & a-1 \\ a-1 & 2(a-1) \end{pmatrix}.$$

($i$) When $a > 1$, $B$ is a positive definite matrix and $x^*$ is asymptotically stable, and $p^* = \frac{1}{3}R + \frac{1}{3}S + \frac{1}{3}P$ is an ESS. All the inner orbits converge to $x^*$.
($ii$) When $a = 1$, $B$ is a positive semi-definite matrix and $B = 0$, $x^*$is Lyapunov stable, but not asymptotically stable, and $\prod_{i=1}^{3} x_i^{\frac{1}{3}} \equiv c$, that is: $x_1 x_2 x_3 \equiv C$.
($iii$) When $a < 1$, $B$ is a negative definite matrix and $x^*$ is unstable, all the orbits except $x \equiv x^*$ will tend to $BdS_3$.

Obviously, most of the classical game literature are discussed base on the extension of the Rock-Scissors-Paper (RSP) game, which is a simple case of N=3 (Hofbauer & Sigmund 1998; Sigmund 2010; Nowak 2006). The objective of this study is to extend it to any $N \geq 2$ cases.

## Other examples

The matrix $B$ has five possibilities: positive definite matrix, positive semi-definite matrix, negative definite matrix, negative semi-definite matrix and indefinite matrix. In this study, only specific theorems and conclusions are given for the first three cases. Let's discuss the last two cases (negative semi-definite matrix and indefinite matrix):

($i$) When $N = 3$ and $A = \begin{pmatrix} 0 & 3 & -1 \\ 3 & 0 & -1 \\ 1 & 1 & 0 \end{pmatrix}$, there is a unique fixed point of equation in $Int\mathrm{S}_3$, that is,

$$x^* = \left(\frac{1}{3}, \frac{1}{3}, \frac{1}{3}\right)$$

which is a saddle point. In this case, $B = \begin{pmatrix} 0 & -6 \\ -6 & 0 \end{pmatrix}$ and $B$ is a negative semi-matrix. The corresponding phase diagram of the equation has been discussed in Figure 14 of Bomze's report (Bomze 1983), where some orbits tend to $x^*$, while some orbits tend to $BdS_3$.

$(ii)$ When $N = 3$ and $A = \begin{pmatrix} 0 & 6 & -4 \\ -3 & 0 & 5 \\ -1 & 3 & 0 \end{pmatrix}$, there is a unique fixed point of equation in $IntS_3$, that is,

$$x^* = \left(\frac{1}{3}, \frac{1}{3}, \frac{1}{3}\right)$$

which is locally asymptotically stable. At the time, $B = \begin{pmatrix} -10 & 0 \\ 0 & 16 \end{pmatrix}$ is an indefinite matrix. The corresponding phase diagram of the equation has been discussed in Figure 12 of Bomze's report (Bomze 1983; Hofbauer& Sigmund 1998), where some orbits tend to $x^*$, while some orbits tend to the $BdS_3$.

These results suggest that further research is needed when *B* is a negative semi-definite matrix or indefinite matrix.